\documentclass[prb,twocolumn,showpacs,amsmath,amssymb,superscriptaddress,floatfix]{revtex4}
\usepackage{balance}
\usepackage{graphicx}
\usepackage{dcolumn}
\usepackage{bm}

\begin{document}

\title{Low-lying excited states quantum entanglement and continuous quantum phase transitions: The criticality of a one-dimensional deconfined critical point}

\author{Yan-Chao Li}
\email{ycli@ucas.ac.cn}
\author{Yuan-Hang Zhou}
\author{Yuan Zhang}
\affiliation{ \textit Center of Materials Science and
Optoelectronics Engineering, College of Materials Science and
Opto-Electronic Technology, University of Chinese Academy of
Sciences, Beijing 100049, China}
\author{ Hai-Qing Lin }
\affiliation{\textit Zhejiang Institute of Modern Physics and School
of Physics, Zhejiang University, Hangzhou 310027, China}
\date{ \today }

\begin{abstract}
From the perspective of low-lying excited states, we study the
deconfined quantum critical point (DQCP) in a one-dimensional
quantum spin chain by means of the entanglement entropy and
fidelity. Our results show that there is a close connection between
the reconstruction of low-lying excitation spectra and the DQCP. The
precise position of the critical point and its continuous nature is
indicated by the singular behavior of the entanglement and fidelity
of the first-excited state. Furthermore, compared with the
Berezinskii-Kosterlitz-Thouless type phase transitions, which also
go beyond the scope of Landau-Ginzburg-Wilson paradigm, we attempt
to reveal the essence of different types of symmetries on both sides
of the DQPT from different manifestations of entanglement
singularity.

\end{abstract}

\maketitle

\section{\label{sec:level1} INTRODUCTION }
Quantum phase transitions (QPTs), which are purely driven by quantum
fluctuations, play a significant role in understanding the quantum
mechanism behind many novel physical phenomena and are one of the
central focuses in modern condensed matter
physics~\cite{SSA11,Wen19}. Traditionally, QPTs are well described
by local order parameters and symmetry breaking theory within the
Landau-Ginzburg-Wilson (LGW) paradigm. However, in recent years,
researchers have found some new QPTs that cannot be classified into
the category of LGW both in theory and experiments, such as
topological QPTs~\cite{Tsui1982,Hasan2010,Qi2011,Hu2020} and
deconfined quantum critical points
(DQCPs)~\cite{Senthil2004,SenthilB2004,Shao2016,Cui2023}. To
incorporate these novel types of phase transitions into a unified
theory, it is imperative to pursue fresh breakthroughs.  As a unique
characteristic of quantum systems, quantum entanglement has made
important progress in the study of quantum phase
transitions~\cite{AOS02,GSJ04,Legeza06,Berkovits2015,Yang2021}.
Utilizing this pure quantum quantity to illuminate quantum phenomena
is regarded as the linchpin in overcoming this challenge.

The DQCP was originally introduced in two-dimensional (2D) quantum
Heisenberg magnets, where a continuous QPT occurs between two
quantum states with irrelevant
symmetries~\cite{Senthil2004,SenthilB2004}. According to the
traditional LGW phase transition theory, due to the irrelevant
symmetries, the transition here should not be continuous, but rather
a first-order phase transition or occur at two separate points. This
novel phenomenon has attracted a significant amount of theoretical
interest~\cite{Sandvik2010,Chen2013,Shao2016,Roberts2019,Huang2019}.
Recently, the experimental demonstration of the existence of the
field-induced DQCP at low temperatures through high-pressure nuclear
magnetic resonance measurements on layered quantum magnet
$SrCu_2(BO_3)_2$ has further piqued researchers' enthusiasm for DQCP
studies~\cite{Cui2023}. However, the nature properties of the phase
transition is still under debate, and the controversial for
continuous or weakly-first order still remains.

Existing theoretical studies have primarily focused on the
non-analytic behavior of ground state properties at the transition
points \cite{Huang2019,Luo2019,Yang2021}. However, for continuous
phase transitions, it is widely recognized that they are not solely
determined by ground state properties, but also significantly
influenced by the interplay between the ground state and low-lying
excited states \cite{SSA11,TGS03,GSJ07}. It has even been proposed
that the quantum phase transition (QPT) is primarily driven by the
reconstruction of the excitation energy spectra \cite{TGS03}.
Furthermore, in the $J_1-J_2$  model, the continuous QPT can be more
effectively captured by the fidelity of the first excited state
rather than the ground state \cite{CS07}. Consequently, it is
imperative to delve into the nature of this phase transition from
the perspectives of low-lying excited states and quantum
entanglement.

Additionally, the Berezinskii-Kosterlitz-Thouless (BKT) type QPT was
originally introduced to describe the proliferation of topological
defects in the two-dimensional (2D) XY spin
model~\cite{Berezinskii1971,Kosterlitz1973}. As there is no genuine
long-range order, topological phase transitions cannot be
characterized by local order parameters and fall outside the LGW
paradigm of symmetry breaking. In one-dimensional (1D) quantum
systems, the continuous QPT with infinite order and lacking a local
order parameter is also known as the BKT-type QPT. Due to the
long-range correlation length and exponentially close gap at the
transition point \cite{Itzykson1989}, numerical investigations of
the BKT phase transition from pure ground-state properties pose a
significant challenge. This is because extremely large system sizes
are required to avoid severe finite-size effects
\cite{GSJ07,YWL07,CS08,Sun2015,LYC22}.


In this study, we focus on exploring the DQCP by examining quantum
entanglement in the low-lying excited states. Our goal is to conduct
a comparative analysis of the behaviors exhibited by excited state
quantum entanglement across BKT-type quantum phase transitions. This
analysis aims to expose the distinctions and connections between two
types of QPTs that fall outside the theoretical framework of LGW
symmetry breaking, despite their distinct origins. Furthermore, we
aim to investigate their association with quantum entanglement and
ultimately reveal the fundamental nature of DQPT formation.

\section{\label{sec:level2} Models and method}
The one-dimension analog of DQCP has been constructed in several
systems~\cite{Jiang2019,Sandvik2004,Mudry2019,Weber2020,Ogino2021}.
Among them one of the most widely studied is a one-dimensional spin
chain model~\cite{Jiang2019}. Its Hamiltonian can be written as
\begin{eqnarray}\label{eq:1}
H=\sum_{j=1}^N\left(-J_x\sigma_j^x\sigma_{j+1}^x-J_z\sigma_j^z\sigma_{j+1}^z\right.\nonumber\\
+\left.K_x\sigma_j^x\sigma_{j+2}^x+K_z\sigma_j^z\sigma_{j+2}^z\right),
\end{eqnarray}
where ${\sigma_j^\alpha}$ (${\alpha=x,z}$) are the Pauli matrices on
stie $j$ and $J_\alpha$ and $K_\alpha$ describe the nearest-neighbor
ferromagnetc and nest-nearest-neighbor antiferromagnetc
interactions, respectively. ${N}$ is the number of spins in the
chain. For comparison with the known results and simplicity, we fix
$J_x=1$ and $K_x=K_z=1/2$, leaving the only driving parameter $J_z$.
It has been pointed out by field theory that a direct continuous QPT
occurs between two states with irrelevant symmetries: the spin-z
ordered ferromagnetic phase (zFM) ($J_z\gg1$) with breaking
$\mathbb{Z}_2$ and the VBS phase($J_z\sim1$) with breaking
translational symmetry~\cite{Jiang2019,Luo2019}. Therefore,  it is a
DQCP in analogy with the 2D counterpart mentioned in
Sec.~\ref{sec:level1}. Subsequent numerical calculations further
confirms the continuous QPT conclusion and predict the critical
point locates at $J_z^c\approx 1.465$ through finite-size scaling
analysis of the order parameter and quantum entanglement, and
~\cite{Huang2019,Luo2019,Yang2021}.

However, it cannot be excluded that the phase transition may be
weakly first-order due to possible deviations in finite-size scaling
analysis or the accuracy of numerical methods. In fact, a tiny
discontinuous jump in the order parameter near the critical point
has been observed, although the authors attributed it to an artifact
of the numerical matrix product state method
~\cite{Roberts2019,Huang2019}. Given that continuous quantum phase
transitions are often associated with excited states, as discussed
in Sec.~\ref{sec:level1}, we approach this issue from the
perspective of low-lying excited states, combining quantum
information approaches.

For the BKT-type QPT, we first consider the 1D spin XXZ model. The
Hamiltonian is defined as follows:
\begin{eqnarray}\label{eq:2}
H=\sum_j^N\sigma _{j}^{x}\sigma _{j+1}^{x}+\sigma _{j}^{y}\sigma
_{j+1}^{y}+\Delta \sigma _{j}^{z}\sigma _{j+1}^{z},
\end{eqnarray}
where ${\Delta}$ describes the anisotropy of the spin-spin
interaction on the z direction, and ${\sigma_j^\alpha}$ and $N$ have
the same meaning in Eq.~(\ref{eq:1}). It is well
known~\cite{Gogolin1999,Venuti2007,CS08} that in the regime
$-1<\Delta<1$ the model is in a critical phase displaying gapless
excitations and power low correlations. For $\Delta>1$ the model
enters a phase with Ising-like antiferromagnetic phase and a nonzero
gap. The isotropic antiferromagnetic point  $\Delta=1$ is a BKT
transition point, which is described by a divergent correlation
length but without true long-range order, and is not easy to be
detected by quantum information and finite-size scaling
approaches~\cite{CS08,Venuti2007,Sun2015,LYC22}.

Another BKT-type QPT that is further considered is the transition at
$J_2/J_1\approx0.241$ of the $J_1-J_2$ model. The Hamiltonian reads
as
\begin{eqnarray}\label{eq:3}
H=\sum^{N}_{j=1}\left(\sigma_{j}\cdot \sigma_{j+1}+\lambda
\sigma_{j}\cdot\sigma_{j+2}\right), \label{eq:3}
\end{eqnarray}
where $\lambda$ describes the ratio between the
next-nearest-neighbor (NNN) interaction $J_2$ and the
nearest-neighbor interaction $J_1$. It is well known there is a
BKT-type quantum phase transition at
$\lambda_c\approx0.241$~\cite{CS07}. For $\lambda<\lambda_c$, it is
a gapless spin fluid or Luttinger liquid phase. As
$\lambda>\lambda_c$, the ground state changes into a spin-gapped
dimerized phase~\cite{KOK92,Castilla1995}.
Similar to the study of DQCP, we investigate the relationship
between excited states and phase transitions, emphasizing the link
between these transitions and the entanglement of excited states.
Furthermore, we conduct a comparative analysis with the case of
DQPT.

The concept of entanglement has been successfully used in detecting
QPTs in various
systems\cite{AOS02,Werlang2010,Berkovits2015,Yang2021}. Here, we
adopt the entanglement of formation (EOF) as the detector, which is
defined as~\cite{Werlang2010}
\begin{align}
E_f(\rho_{AB})=&-f(C_{\rho_{AB}})\log_2f(C_{\rho_{AB}})\nonumber\\
&-\left[1-f(C_{\rho_{AB}})\right]\log_2\left[1-f(C_{\rho_{AB}})\right],\label{eq:5}
\end{align}
where ${\rho_{AB}}$ is the reduced density matrix of two neighboring
sites $A$ and $B$ in the spin chain.
$f\left(C_{\rho_{AB}}\right)=(1+\sqrt{1-C_{\rho_{AB}}^2})/2$ is a
monotonically increasing function of the concurrence
$C_{\rho_{AB}}$\cite{WKW98}.
$C_{\rho_{AB}}=\max\{0,\lambda_1-\lambda_2-\lambda_3-\lambda_4\}$,
where $\lambda_1$, $\lambda_2$, $\lambda_3$, and $\lambda_4$ are the
square roots of the eigenvalues of $\rho_{AB}\tilde{\rho}_{AB}$ in
descending order.
$\tilde{\rho}_{AB}=(\sigma_A^y\otimes\sigma_B^y)\rho_{AB}^\ast(\sigma_A^y\otimes\sigma_B^y)$
is the time-reversed matrix of $\rho_{AB}$. $\rho_{AB}^\ast$ is the
complex conjugation of $\rho_{AB}$ and $\sigma^y$ is the $y$
component of Pauli operator. We will analyze the reduced density
matrices ${\rho_{AB}}$ formed by the ground state wave function
$\psi_g$ and the first-excited state wave function $\psi_e$,
respectively. The resulting $E_f(\rho_{AB})$ are accordingly denoted
as $E_f^g$ and $E_f^e$.

Another widely utilized detector for QPT study from the field of
quantum information theory is the quantum fidelity (QF). It is
defined as the overlap between two quantum states ${\psi(\lambda)}$
and ${\psi(\lambda+\delta})$ as
$F(\lambda,\delta)=|\langle\psi(\lambda)|\psi(\lambda+\delta)\rangle|$,~\cite{PZA06,CS07,WLY07,GSJ10,Sun2015}.
Here, $\lambda$ is a driving parameter, while $\delta$  is a small
quantity. If $\lambda$ and $\lambda+\delta$ fall into different
quantum states due to the vastly distinct geometrical structures of
these states in Hilbert space, the value of $F(\lambda,\delta)$ will
deviate from 1, making it a suitable indicator for QPT. Similar to
the case of $E_f$, if the first excited quantum state is used to
calculate F, it is labeled as $F_e$. We will also conduct
comparative analysis on EOF results using QF.

We use the numerical exact diagonalization (ED) techniques to
simulate the two spin systems up to $N=24$. To reduce the influence
of the boundary, periodic boundary conditions are considered for the
EOF and QF calculations.

\section*{\label{sec:level3} RESULTS AND DISCUSSIONS}
\begin{figure}
\includegraphics[width=8.5cm]{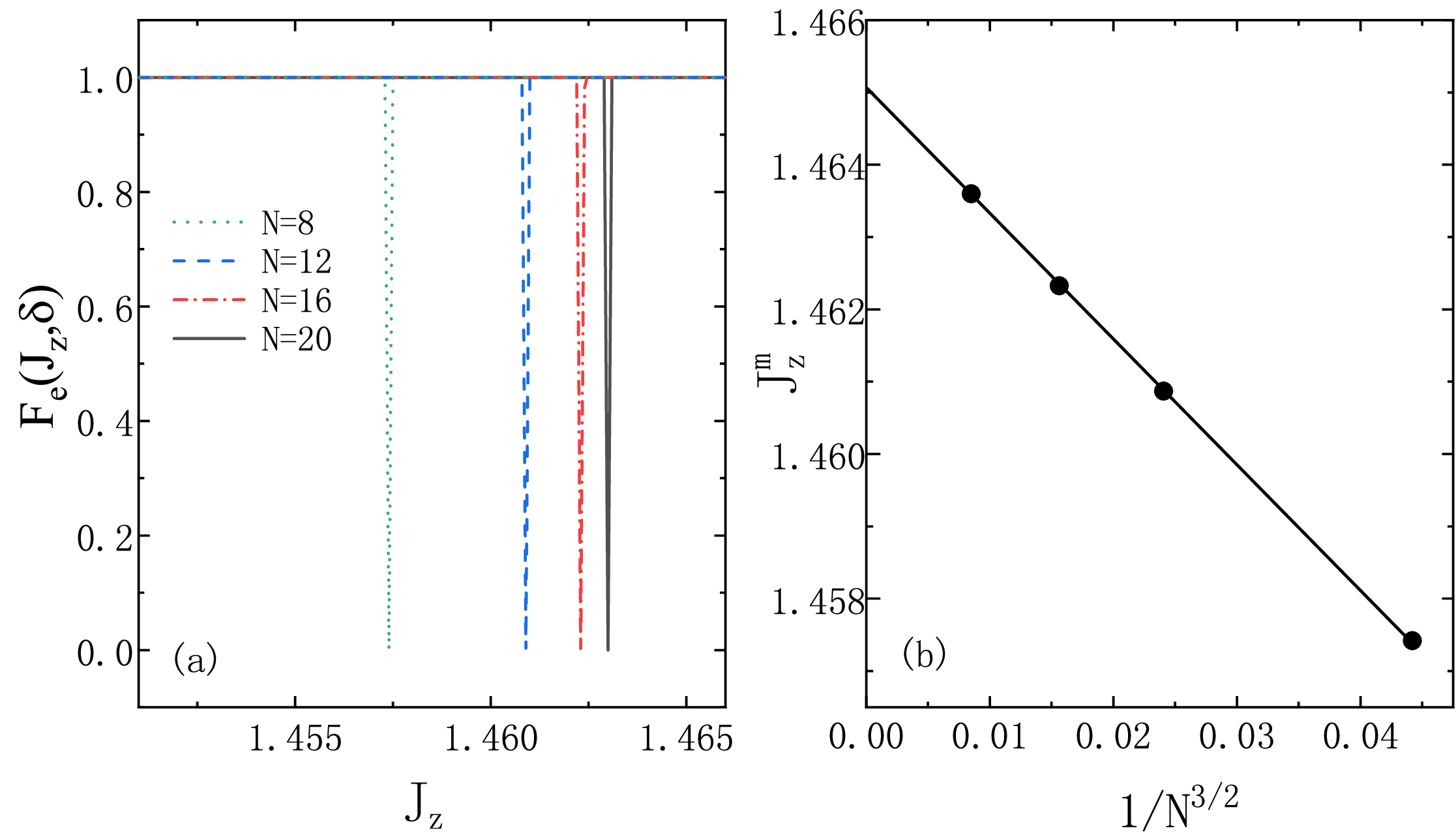}
\caption{\label{fig1} (Color online) (a) Fidelity of the first
excited state $F_1(J_z,\delta)$ with $\delta=1\times10^{-4} $as a
function of $J_z$ under different system size $N$ for DQPT model.
(b) Linear finite-size scaling of the extrema $J_z^m$ (black dots)
of $F_1(J_z,\delta)$ in (a). When $N\rightarrow\infty$, the
extrapolated critical value is $J_z^c=1.4645(1)$.}
\end{figure}

\begin{figure}
\includegraphics[width=6cm]{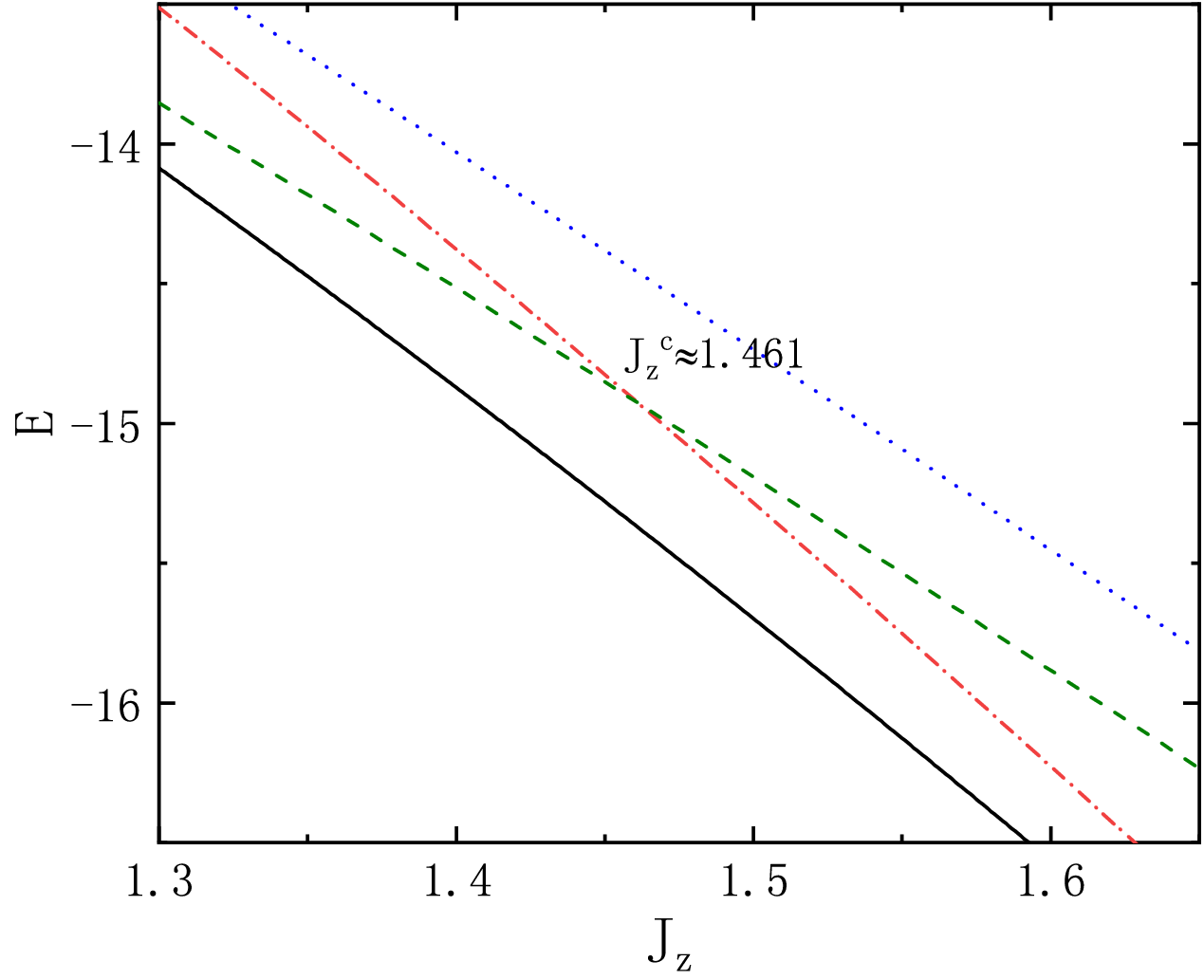}
\caption{\label{fig2} (Color online) Energy spectrum of the DQPT
model with $N=12$. There is a level crossing at $J_z\approx1.461$
for the first excited state.}
\end{figure}

For the 1D DQPT model, as expected, due to the continuity of the
ground state at the critical point, the ground state fidelity $F_g$
does not exhibit any singularity. However, the first excited state
fidelity $F_e$ with $\delta=1.0\times10-4$ shows a sudden drop and
its position $J_z^m$ moves to the critical point $J_z^c$ as $N$
increases (see Fig.~\ref{fig1}). It is clear that there is a
size-scaling behavior for $J_z^m$. To further confirm this
conclusion, we do the finite-size scaling analysis for $J_z^m$. As
shown in Fig.~\ref{fig2}, the system-size dependent $J_z^m$ linearly
scales as $1/N^{3/2}$ with error less than $\pm 1\times10^{-5}$. The
extrapolated value to $N=\infty$ is $J_z^c\approx1.4645(1)$, which
consists with the best estimates results in
Refs.~\onlinecite{Huang2019,Luo2019,Yang2021} very well.

Now we can confirm that the DQPT can be characterized by the first
excited state fidelity. This is quite similar to the case of the
$J_1-J_2$ model in Ref.~\onlinecite{CS07}, where the continue QPT at
$J_2\approx0.241$ can only be detected by the first excited state
fidelity, but not the ground state fidelity. This means that the
DQPT here should also be caused by the level crossing of
low-excitation spectrum of the system. It is confirmed by the energy
spectrums of the system for $N=12$ as shown in Fig.~\ref{fig2}, one
can clearly see that the ground sate is continuous while the first
excited state is level crossing at $J_z\approx1.461$. As pointed out
in Ref.~\onlinecite{TGS03}, this origin from the energy level
reorganization of the low-energy excited state reflects the
continuous nature of the DQPT, from which we can rule out the
possibility of this DQPT being a weakly first-order phase
transition.

\begin{figure}[b]
\includegraphics[width=8cm]{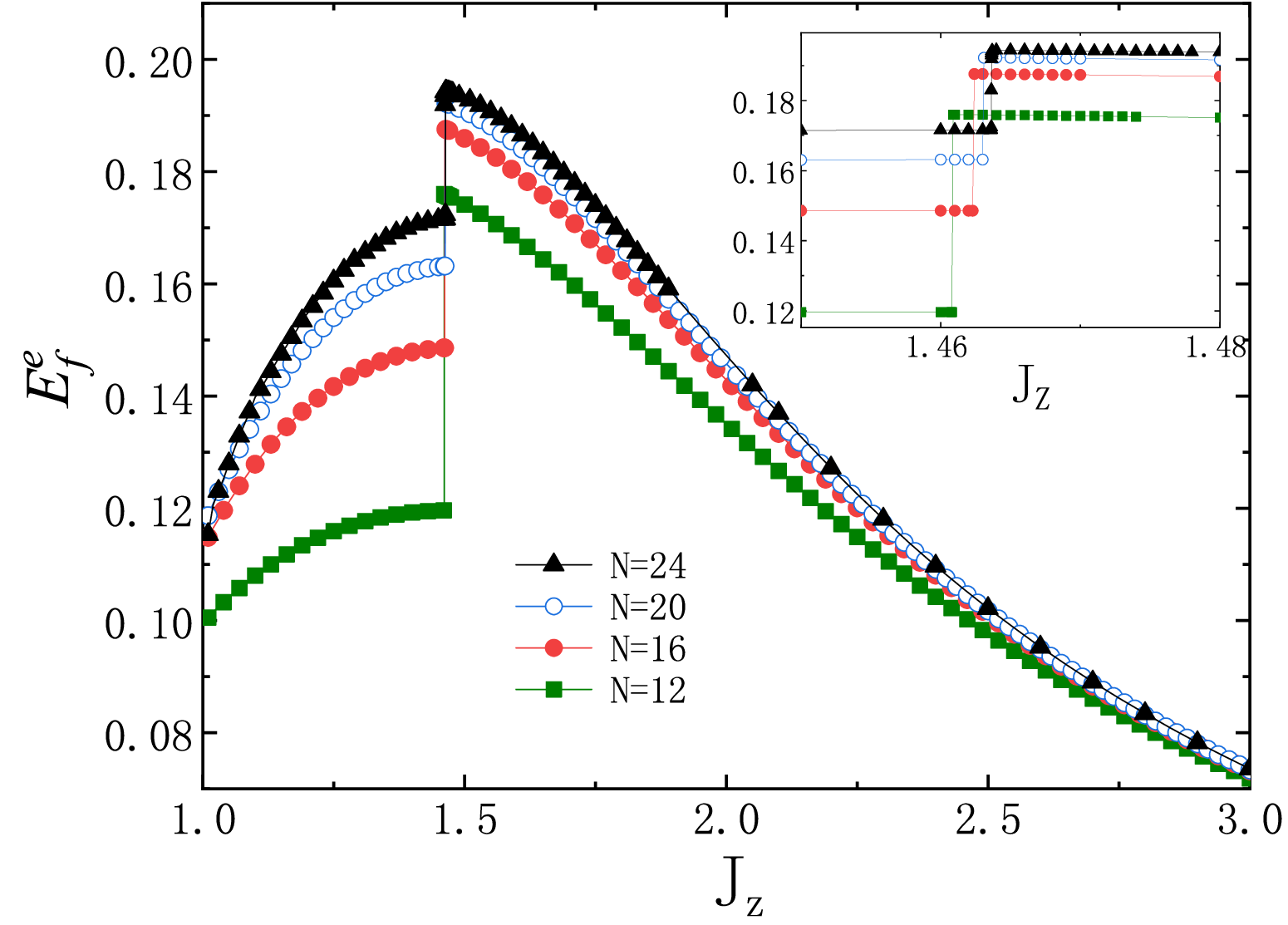}
\caption{\label{fig3}(Color online) (a) First excited state
entanglement of formation $E_f^e$ as a function of $J_z$ under
different $N$. The inset is an enlarged view of the curves near the
jump for better clarity.}
\end{figure}

Now that both the DQPT and BKT-type transitions belong to the
continuous category, and the dropping feature of the fidelity
obviously cannot reflect their difference. How to identify their
respective characteristic is still a problem. To deal with this
issue we turn to the quantum entanglement. The first excited state
entanglement of formation $E_f^e$ as a function of $J_z$ under
different system size $N$ is plotted in Fig.~\ref{fig3}. A very
remarkable phenomenon is the jump behavior of the curve near the
phase transition point.

We discovered that the position of this jump point and its changes
in relation to $N$ (see the local enlarged view as an inset at the
bottom of Fig.~\ref{fig3}) align perfectly with the critical point
behavior described by $F_1$. This indicates that EOF in the first
excited state can clearly reflect this DQPT. Furthermore, $E_f^e$
exhibits an intriguing characteristic: it attains a maximum on both
sides of the jump point. This finding suggests that if the jump
point indeed marks a critical point, then both quantum states reach
their maximum entanglement at the phase transition point.

\begin{figure}[t]
\includegraphics[width=7cm]{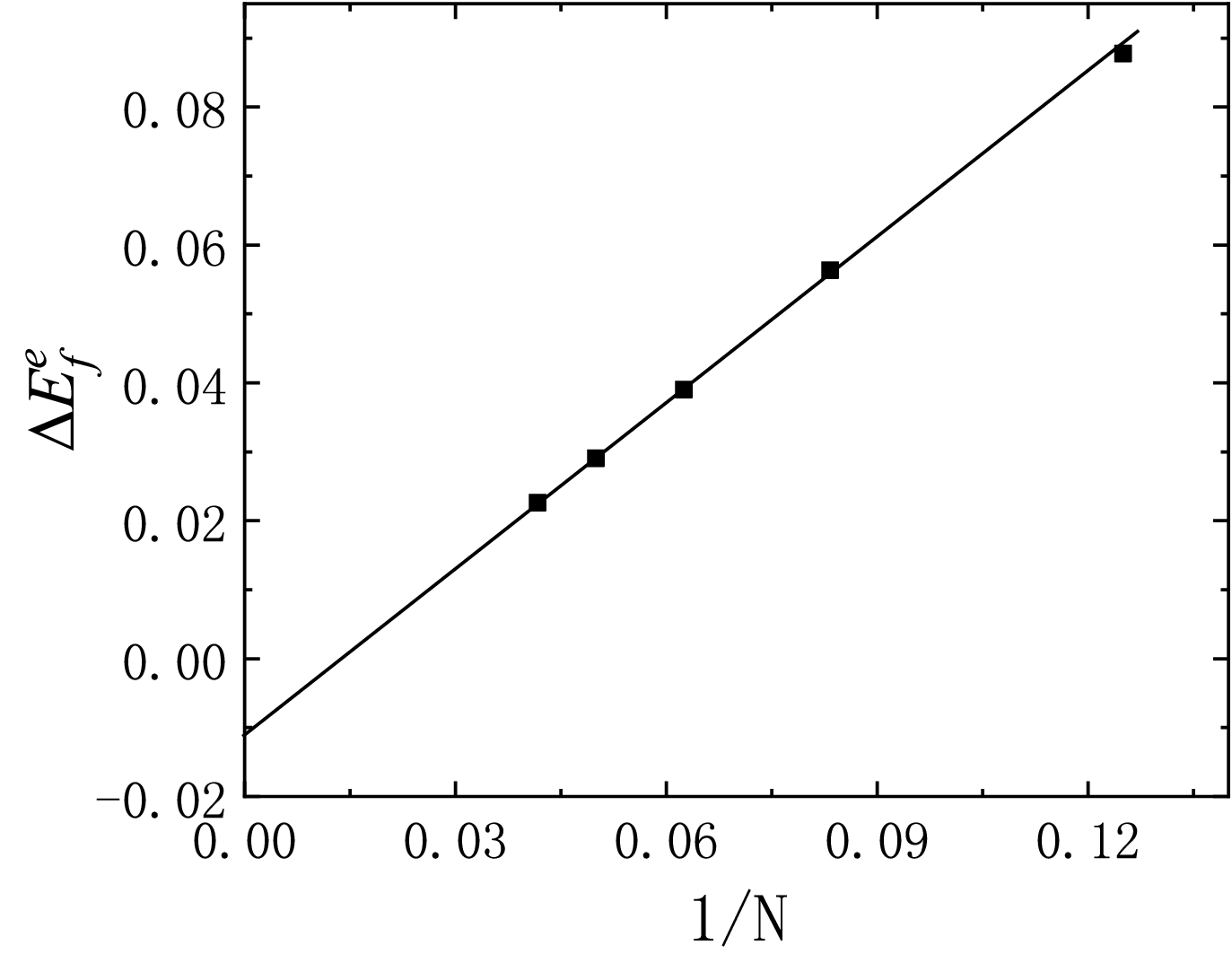}
\caption{\label{fig4}(Color online) (a) Finite-size scaling of the
entanglement gap $\Delta E_f^e$ versus $1/N$. The black line is a
linear fit, which indicates that the gap will not disappear in the
thermodynamic limit.}
\end{figure}

We think this phenomenon completely reflect the feature of the DQPT
and can be explain as follows:  The degree of symmetries of the VBS
and zFM phases besides the critical point becomes higher as they
approach the critical point, and the degree of entanglement is
usually related to the symmetries of the system for a specific
quantum state, leading to the maximal entanglement. This is quite
similar to the entanglement behavior for the XXZ model pointed out
in Ref.~\onlinecite{GSJ07}. However, there are also differences: As
mentioned in Sec.~\ref{sec:level2}, the quantum states located on
different sides of the DQPT exhibit complete breakdown of symmetries
that are unrelated to each other. Consequently, for the DQPT system,
the entanglement difference between $E_f^e$ on both sides of the
phase transition point is substantial, featuring a distinct jump at
the phase transition point. Furthermore, the entanglement gap
$\Delta E_f^e=E_{f,R}^e-E_{f,L}^e$, where  $E_{f,R}^e$ and
$E_{f,L}^e$ are the  $E_{f}^e$ on the right and left sides of the
jump point, respectively, follows a linear scaling with 1/N as shown
in Fig.~\ref{fig4}. It appears that the jump will not disappear even
in the thermodynamic limit, reflecting the distinct quantum features
of the quantum state beyond the critical point. Nevertheless, it's
worth noting that the symmetries of the quantum states on both sides
of the BKT transition point in the XXZ model belong to the same
category. Both sides undergo a transition from $SU(2)$ to $SU_q(2)$
at the phase transition point~\cite{GSJ07}, with the only difference
being in the dominant direction. Therefore, we infer that the
entanglement of the first excited state for the XXZ model should be
continuous at the phase transition point.

\begin{figure}[t]
\includegraphics[width=8.5cm]{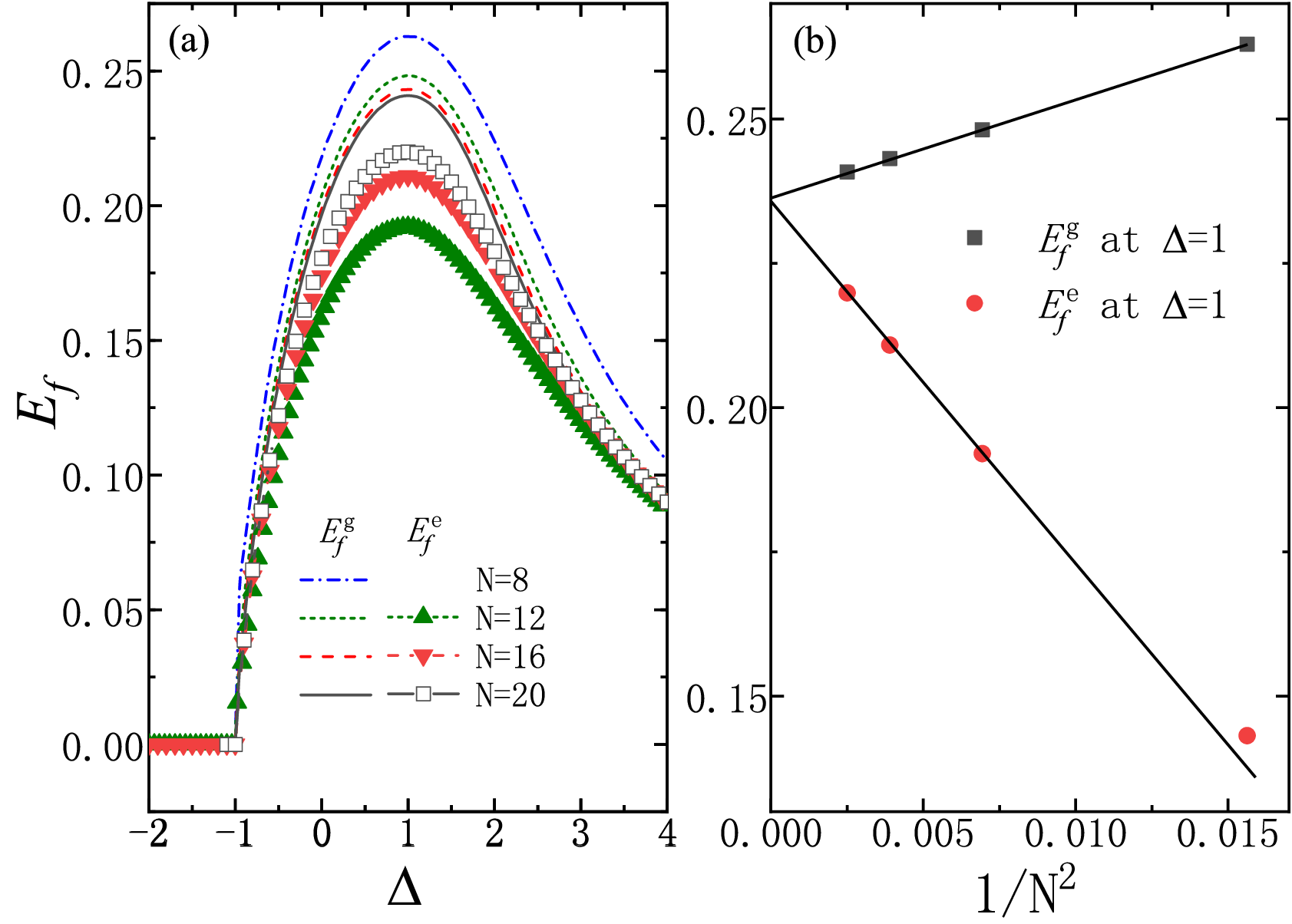}
\caption{\label{fig5}(Color online) (a) First excited state EOF
$E_f^e$ (with symbols) and ground-state EOF $E_f^e$ (without
symbols) as a function of $\Delta$ for different system size $N$ in
the XXZ model. (b) Linear finite-size scaling of $E_f^g$ (black
squares) and $E_f^e$ (red dots) at the critical point $\Delta=1$. In
the thermodynamic limit $E_f^e$ and $E_f^g$ will approach the same
value.}
\end{figure}

Figure~\ref{fig5} shows the behavior of the first excited sate
entanglement $E_f^e$ as a function of $\Delta$ under different
system sizes for the XXZ model. Although there is a level crossing
in the first excited state at $\Delta=1$~\cite{GSJ07}, $E_f^e$ is
indeed continuous and shows a round peak at the transition point as
we expected. It has been pointed out that the ground state
entanglement show a maximum behavior and can be used to detect the
BKT-type transition~\cite{Wang2015}, and it is explained that it is
the influence from the low-lying excited state to the ground state
cause the maximum behavior. Here, we note that the first excited
state, as described in Ref.~\onlinecite{TGS03}, lies at the core of
this ongoing transition. Entanglement on this state can serve as a
straightforward indicator of the transition's occurrence. In fact,
as shown in Fig.~\ref{fig5}(a), the peak values for the ground and
first excited states entanglement decrease and increase,
respectively, as N increases. It appears that they will converge
into a single curve. The finite-size scaling behavior of $E_f^e$ and
$E_f^g$ at the critical point $\Delta=1$ supports this observation
(see Fig.~\ref{fig5}(b)): both of them follow a linear scaling
behavior with $1/N^2$. As $N$ approaches infinity, both of them
approach the same value of approximately $0.238$, indicating that
the entanglement behavior of the ground and first excited states
will tend to agree in the thermodynamic limit. This fully reflects
the important role played by the low-excited states in the formation
of such continuous quantum phase transitions.

\begin{figure}[t]
\includegraphics[width=8.5cm]{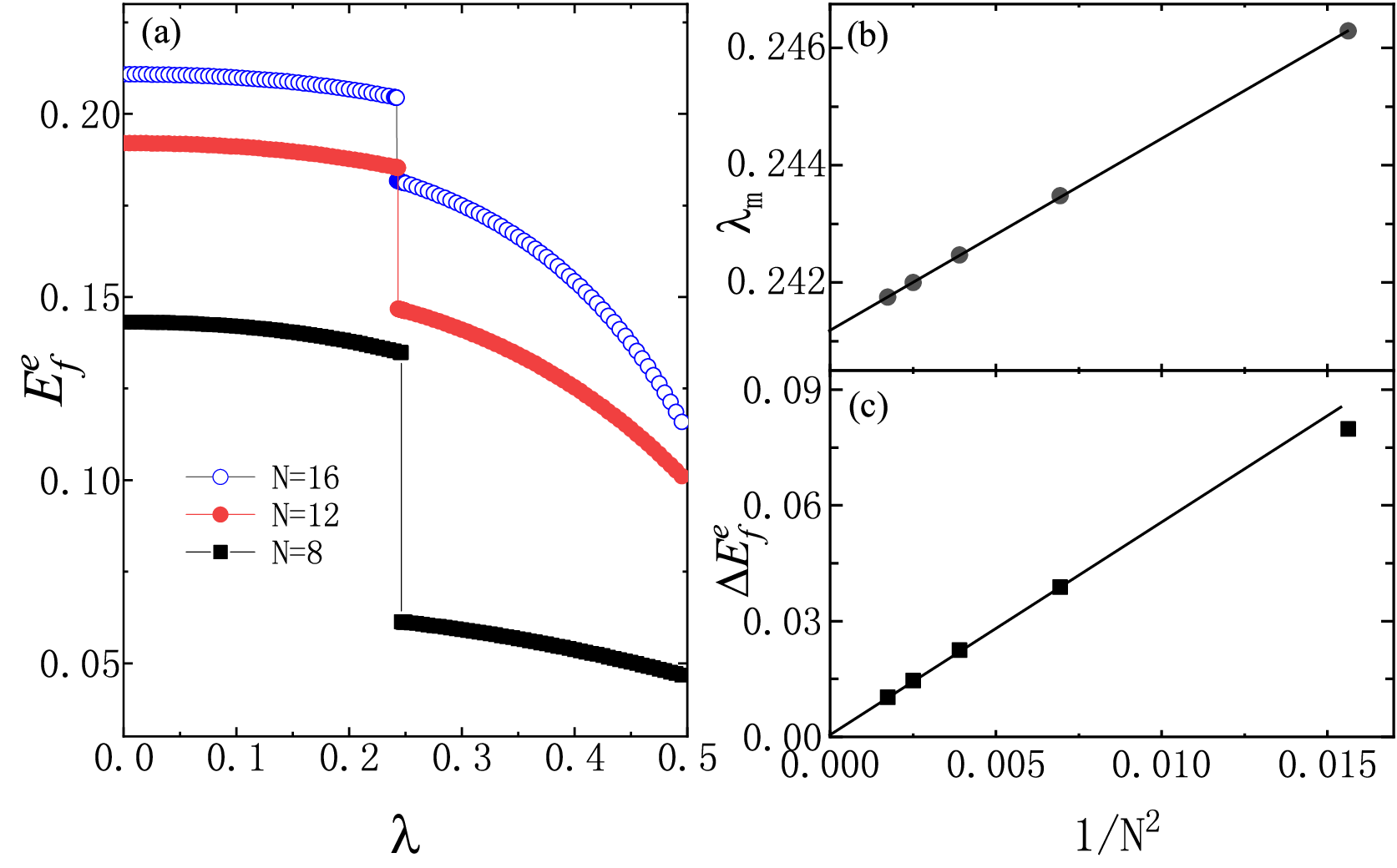}
\caption{\label{fig6}(Color online) (a) First excited state EOF
$E_f^e$ as a function of $\lambda$ for different system size $N$ in
the $J_1-J_2$ model. Linear finite-size scalings of (b) the jump
position $\lambda_m$ and (c) the entanglement gap $\Delta E_f^e$
beside the jump as a function of $1/N^2$. The critical point is
extrapolated to be $\lambda_c=0.2412(0)$, and the entanglement gap
disappears in the thermodynamic limit.}
\end{figure}

We then consider the BKT-type QPT in the $J_1-J_2$ model for a
further comparison. The first excited state entanglement $E_f^e$
under different $N$ is shown in Fig.~\ref{fig6}(a). Although there
is no maximum behavior, there is a jump that moves towards the
critical point as N increases. The jump position $\lambda_m$ scales
linearly with $1/N^2$ as shown in Fig.~\ref{fig6}(b), indicating the
critical point $\lambda_c\approx0.2412(0)$ in the limit of
$N\rightarrow\infty$. It is evident that the scaling behavior of the
jump in $E_f^e$  and the sudden drop in fidelity (reported in
Ref.~\onlinecite{CS07}) are consistent. This consistency illustrates
that the jump observed here also reflects the level crossing in the
first excited state and can be utilized to detect the continuous
QPT.


The behavior of $E_f^e$ differs from that in both the DQPT and XXZ
models: there is no maximum phenomenon, and although it exhibits a
jump similar to that in the DQPT model, it will disappear in the
thermodynamic limit, as shown in Fig.~\ref{fig6}(c). The linear
scaling relationship between the gap before and after the jump,
$\Delta E_f^e$, and $1/N^2$ indicates that $\Delta E_f^e$ disappears
as $N$ approaches infinity. We believe that the main reason for this
is still related to the specific symmetry in the system. Unlike the
DQPT with irrelevant symmetry breaking and the XXZ model with the
same kind of symmetry breaking, only one-side symmetry breaking
occurs for the dimerized state. As the driving parameter $\lambda$
increases, the spin frustrated effect is enhanced, and the
translational symmetry is gradually decreasing. However, due to the
fact that these two phases belong to the same symmetry category,
there is no jump for $E_f^e$ in the thermodynamic limit, just as it
is observed in the XXZ model. Furthermore, for a finite-size system,
$E_f^e$ does not exhibit a jump at the BKT phase transition of the
XXZ model. We suggest that this phase transition is solely caused by
anisotropy, which is not linked to system size like translation
symmetry, but rather depends solely on spin direction. Consequently,
there is no size scaling behavior and no jump behavior.


\section{\label{sec:level4} summary}
In summary, we have delved into the DQPT and BKT-type QPTs by
examining the entanglement of low-lying excited states. Our findings
indicate that the DQPT is triggered by the level crossing of the
first excited state, much like the BKT-type QPTs, confirming its
continuous nature. Through finite-size scaling analysis, we have
identified the first excited state EOF $E_f^e$ as a valuable tool
for detecting these continuous QPTs. Furthermore, we observed that
the curvature of $E_f^e$  near the critical point offers insights
into the symmetry properties of the phases beyond this juncture.
Notably, irrelevant symmetry breaking results in a jump in $E_f^e$
for the DQPT model, whereas the BKT-type QPT exhibits no such jump
behavior in the thermodynamic limit, as states within the same
symmetry category remain unaltered.

\begin{acknowledgments}
We acknowledge financial support from the National Natural Science
Foundation of China (Grants No. 12074376 and No. 52072365), the
Beijing Municipal Natural Science Foundation (Grants No.1222027),
and the NSAF (Grants No. U1930402).
\end{acknowledgments}


\end{document}